\documentclass{ecai} 

\usepackage{latexsym}
\usepackage{amssymb}
\usepackage{amsmath}
\usepackage{amsthm}
\usepackage{booktabs}
\usepackage{enumitem}
\usepackage{graphicx}
\usepackage{color}
\usepackage{listings}

\usepackage{booktabs}       
\usepackage{amsthm}
\usepackage{amsmath}
\usepackage{dsfont}
\usepackage{subcaption}
\usepackage{float}
\usepackage[hidelinks]{hyperref}

\usepackage{siunitx}

\definecolor{peachysunset}{RGB}{255,183,98}

\usepackage[textsize=small]{todonotes}

\usepackage{soul,xcolor}
\setul{-2ex}{2ex}

\usepackage{dblfloatfix}

\newcommand{\BibTeX}{B\kern-.05em{\sc i\kern-.025em b}\kern-.08em\TeX}

\begin{document}
\begin{frontmatter}
    \paperid{42} 
    \title{whittlehurst: A Python package implementing \\ Whittle's likelihood estimation of the Hurst exponent}
    
    \author[A]{\fnms{Bálint}~\snm{Csanády}\thanks{Corresponding authors: csbalint@protonmail.ch, andras.lukacs@ttk.elte.hu}}
    \author[A,B]{\fnms{Lóránt}~\snm{Nagy}}
    \author[A,\textasteriskcentered]{\fnms{András}~\snm{Lukács}} 
    
    \address[A]{Institute of Mathematics, ELTE Eötvös Loránd University, Budapest, Hungary}
    \address[B]{Alfréd Rényi Institute of Mathematics, Budapest, Hungary}

    
    \begin{abstract}
        This paper presents \texttt{whittlehurst}, a Python package implementing Whittle's likelihood method for estimating the Hurst exponent in fractional Brownian motion (fBm).
        While the theoretical foundations of Whittle’s estimator are well-established, practical and computational considerations are critical for effective use.
        We focus explicitly on assessing our implementation's performance across several numerical approximations of the fractional Gaussian noise (fGn) spectral density, comparing their computational efficiency, accuracy, and consistency across varying input sequence lengths.
        Extensive empirical evaluations show that our implementation achieves state-of-the-art estimation accuracy and computational speed.
        Additionally, we benchmark our method against other popular Hurst exponent estimation techniques on synthetic and real-world data, emphasizing practical considerations that arise when applying these estimators to financial and biomedical data.
    \end{abstract}

\end{frontmatter}

\section{Introduction}
    Introduced by Harold Edwin Hurst \cite{hurst1951long}, the Hurst exponent is a key parameter in characterizing fractional Brownian motion (fBm), and quantifies long-range dependencies in various stochastic processes encountered 
    in financial mathematics, for a comprehensive list of materials see \cite{LongMemory}.
    The concept also finds applications in diverse areas such as climate change \citep{yuan2022impact,franzke2015dynamical}, hydrology \citep{hurst1956problem}, detection of epilepsy \citep{acharya2012application}, DNA sequencing \citep{lopes2006long}, data networks \citep{Willinger2001LongRangeDA}, and in cybersecurity through anomaly detection \citep{li2006change}.
    Accurate and computationally efficient estimation of the Hurst exponent is essential for both theoretical analyses and practical applications.

    Whittle's likelihood estimation provides an efficient, frequency-domain framework for estimating the Hurst exponent by fitting the theoretical spectral density to the empirical periodogram \cite{whittle1953analysis, whittle1953estimation}.
    While Whittle's method itself is classical and thoroughly studied, due to the complexity involved in evaluating the fGn spectral density, practical implementations show great differences.
    
    The central goal of this paper is to present and rigorously evaluate \texttt{whittlehurst}, our Python implementation of Whittle’s likelihood estimator \cite{whittlehurst}, highlighting practical performance aspects such as accuracy, computational speed, and numerical stability.
    The spectral density of fGn is represented by infinite sums that imply computationally heavy calculations, thus our implementation compares four distinct approximation strategies: naïve truncation, Paxson's corrected truncation, reformulation via the Hurwitz zeta function, and a Taylor series approximation \cite{shi2024fractional}.
    
    We conduct comprehensive numerical experiments to determine the most effective spectral density approximation, balancing computational efficiency with estimation accuracy.
    Additionally, we benchmark the performance of our implementation against other widely-used Hurst estimators, such as rescaled range analysis \cite{hurst1951long}, Higuchi’s fractal dimension \cite{higuchi1988approach}, detrended fluctuation analysis \cite{peng1994mosaic}, variogram \cite{gneiting2012estimators}, and time-domain maximum likelihood methods.
    Importantly, these experiments are conducted on independently generated datasets to ensure unbiased assessment of performance.
    
    Moreover, we illustrate practical implications of different estimators through applications on real-world datasets from financial market volatility (S\&P 500, VIX) and EEG signals, providing insights into how data preprocessing steps influence estimation outcomes.
    Our findings confirm that our Python implementation of Whittle's estimator is robust, efficient, and practically reliable, thus offering valuable insights to practitioners requiring accurate and computationally feasible Hurst exponent estimation.

\section{Theoretical background}
    \subsection{Fractional Brownian motion}\label{ssec:fbm-def}
        Fractional Brownian motion, $B_t^H$, is a continuous-time self-similar Gaussian process, characterized by its Hurst exponent $H \in (0,1)$ through its covariance function \cite{mandelbrot1968fractional}:
        $$
          \mathrm{Cov}\bigl(B^H_t,B^H_s\bigr) \;=\; \frac12 \bigl(|t|^{2H} + |s|^{2H} - |t-s|^{2H}\bigr),\;t,s \ge 0.
        $$
        The Hurst exponent controls the long-range dependence of fBm.
        The parameter range $H < 0.5$ results in anti-persistent behavior, and negative correlation of increments, while $H>0.5$ implies persistent trajectories and positively correlated increments.
        The case $H=0.5$ yields standard Brownian motion.
        
        Increments considered on integer time steps form fractional Gaussian noise (fGn), a stationary process exhibiting power-law decay in its autocorrelation function.
        On one hand, when $H<0.5$, fGn displays short memory or negative memory, that is, the infinite sum formed by the lagged auto-correlations is convergent.
        On the other hand, when $H>0.5$, the increment process exhibits long memory, that is, the former infinite sum diverges (see \cite{Giraitis2012}).

    \begin{figure}[H]
        \centering
        \includegraphics[width=0.472\textwidth]{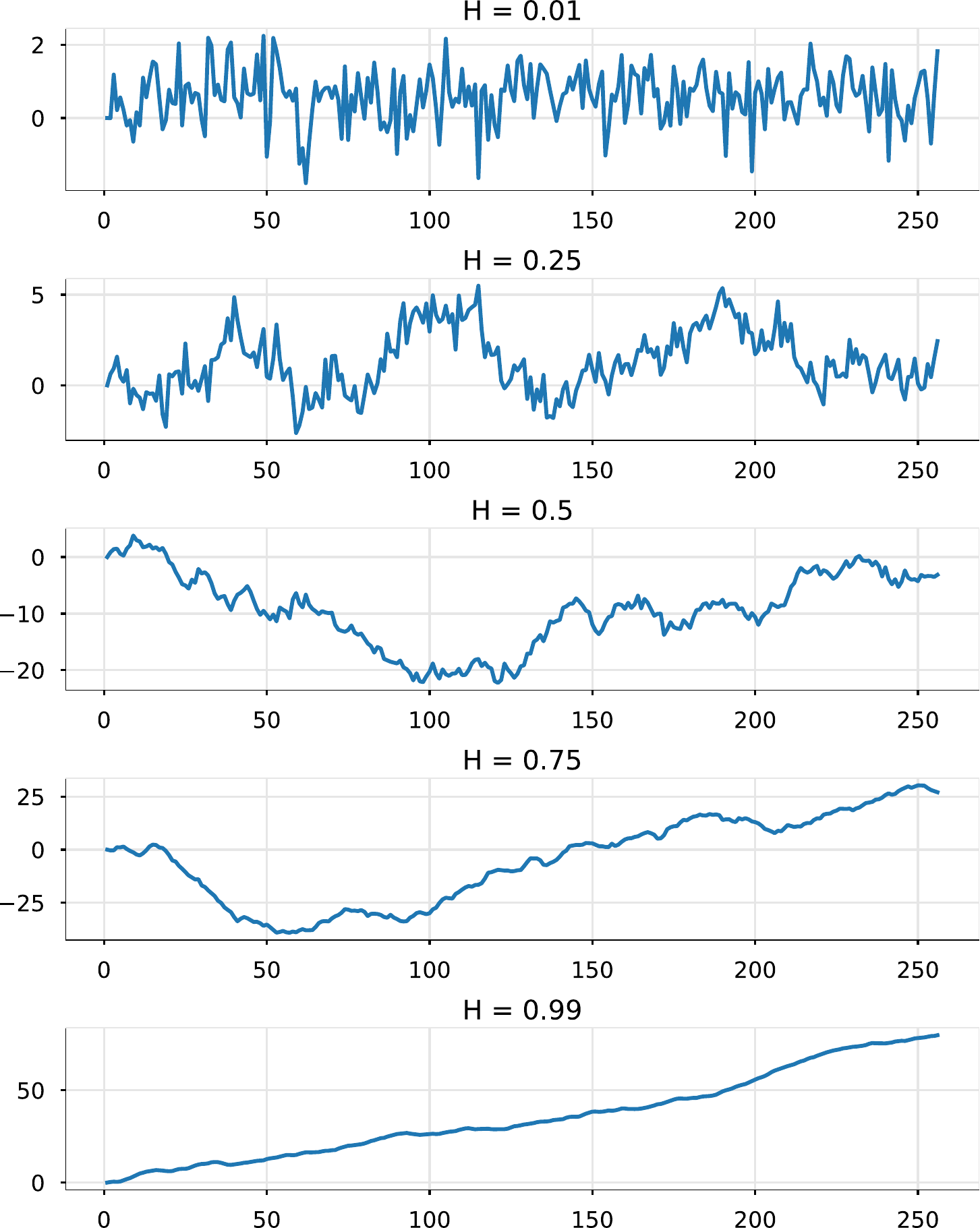}
        \vskip 2mm
        \caption{Example fBm realizations at various $H$ values.}
        \label{fig:fBm_realizations}
    \end{figure}

    \subsection{Whittle's likelihood estimation}
        Spectral densities are key objects in the theory of stationary processes: vaguely speaking, representing the distribution of variance of a process across  different frequencies.
        For fGn, the periodogram provides an empirical estimate of the spectral density.
        Whittle's likelihood estimation \cite{whittle1953analysis, whittle1953estimation} leverages this by fitting the theoretical spectral density to the observed one (the periodogram).
        Maximizing the Whittle likelihood with respect to $H$ yields an estimation procedure in the frequency domain.
        
        The method involves minimizing the Whittle objective function\footnote{Since the method involves finding the minimal argument, constant terms can be omitted.}
        \begin{align*}
            \hat H_\mathbf{y} &:= \arg\min\limits_H \mathcal{L}(H, \mathbf{y}),\\
            \mathcal{L}(H, \mathbf{y}) &= \sum_{j=1}^{\lfloor n/2\rfloor}\frac{I_\mathbf{y}(\lambda_j)}{f_H^*(\lambda_j)},
        \end{align*}
        with $\lambda_j=2\pi j/n$ frequencies, $\mathbf{y} = (y_1, y_2, \dots, y_n)$ observed realization, the normalized theoretical spectral density $f_H^*(\lambda)$, and the periodogram
        $$
        I_\mathbf{y}(\lambda) \cong \left| \operatorname{FFT}\{\mathbf{y}\}(\lambda) \right|^2 = \left| \sum_{t=1}^{n} y_t \, e^{-i \lambda t} \right|^2 .
        $$
        
        We implement the above minimization using the \texttt{fminbound} function from \texttt{scipy} \cite{2020SciPy-NMeth}, that relies on Brent's method \cite{brent2013algorithms}.
        FFT stands for Fast Fourier Transform \cite{fft}, and it computes the 1-D $n$-point discrete Fourier Transform (also imported from \texttt{scipy}).
        Assuming that the theoretical spectral density $f$ can be efficiently computed, Whittle's frequency-domain approach is both computationally efficient and accurate, particularly for large samples \cite{fox1986large, dahlhaus1989efficient}.
        
        For example for the $\operatorname{ARFIMA}(0,H-0.5,0)$ process, the theoretical spectral density can be computed easily as $f_H(\lambda) = \frac{\sigma^2}{2\pi}\cdot(2\sin(\lambda/2))^{1 - 2H}$, see 
        \cite{Giraitis2012}.
        Moreover, the $\operatorname{ARFIMA}$ spectrum is self-normalized, making the implementation of Whittle's method fairly straightforward.
        For fGn however, the spectral density calculation is more challenging.
        
    \subsection{fGn spectral density}\label{subsec_fgn_spect}
        The spectral density of fGn is given by an infinite sum:
        $$
        f_H(\lambda) = \frac{\sigma^2}{\pi}\Gamma(s)\sin(\pi H) (1-\cos\lambda) \sum_{k=-\infty}^{\infty} |2\pi k+\lambda|^{-s},
        $$
        where $s:=2H+1$.
        For Whittle's method, we normalize the spectral density by its geometric mean:
        $$
        f_H^*(\lambda) = \frac{f_H(\lambda)}{\exp\left(\frac{1}{\lfloor n/2\rfloor}\sum\limits_{j=1}^{\lfloor n/2\rfloor} \log\left(f_H(\lambda_j)\right)\right)}.
        $$
        The infinite sum in the spectral density expression makes it difficult to calculate, especially since it must be computed multiple times when minimizing the Whittle objective.
        In the current paper, we present the implementation and comparison of four different methodologies for calculating the spectral density of fractional Gaussian noise (fGn), as discussed in \cite{shi2024fractional}.
        These approaches are briefly summarized below.

        \subsubsection{Approximation by truncation}
            The most straightforward approach is to calculate $f_H(\lambda)$ by truncating the infinite sum (e.g at $K=200$).
            This approach unfortunately yields inaccurate results (especially for smaller values of $H$), and is still computationally intensive.

        \subsubsection{Paxson's approximation}
            Paxson's approximation \cite{paxson1997fast} refines the truncation approach by a correction term.

            $$
            \sum_{k=-\infty}^{\infty} |2\pi k+\lambda|^{-s} \approx \sum_{k=-K}^{K} |2\pi k+\lambda|^{-s} + \frac{1}{2}\Big[a(K,\lambda)+a(K+1,\lambda)\Big],
            $$
            where
            $$
            a(k,\lambda) := \frac{\left(2\pi k+\lambda\right)^{1-s}+\left(2\pi k-\lambda\right)^{1-s}}{4\pi H}.
            $$
            This approach converges much faster compared to the na\"ive truncation, enabling far fewer terms in the calculation (e.g. $K=10$), with better results.

        \subsubsection{Reformulation with the Hurwitz zeta function}
            It is possible to rewrite the infinite sum in terms of the Hurwitz zeta function $\zeta(s,q)=\sum_{j=0}^{\infty}(j+q)^{-s}$:
            $$
            \sum_{k=-\infty}^{\infty} |2\pi k+\lambda|^{-s} = \frac{\zeta\left(s, 1-\frac{\lambda}{2\pi}\right) + \zeta\left(s, \frac{\lambda}{2\pi}\right)}{(2\pi)^{s}}.
            $$
            At first glance, this reformulation may seem pointless: we still have the same infinite sum, now hidden inside $\zeta$.
            However, as we will see in the experiments, it enables us to leverage efficient implementations that approximate $\zeta$ \cite{2020SciPy-NMeth}, and thus yielding a good Whittle implementation in terms of speed-accuracy balance.

        \subsubsection{Taylor series approximation}
            For small frequencies, a Taylor series expansion yields the approximation
            $$
            f_H(\lambda) \approx \frac{\sigma^2}{\pi}\Gamma(2H+1)\sin(\pi H)\, \lambda^{1-2H}.
            $$
            This local approximation is valid as $\lambda\to 0$, so its accuracy is limited to the near-zero frequency region.
            When applied in Whittle's method (which uses a broader range of frequencies), the approximation introduces a bias in the estimation of $H$.

    \subsection{Time-Domain ML estimation}
        The time-domain maximum likelihood (TDML) approach \cite{beran2017statistics, shi2024fractional} estimates the Hurst exponent directly in the time domain by maximizing a likelihood function derived from the process's autocovariance structure.
        Instead of explicitly inverting the full covariance matrix -- an $\mathcal{O}(n^3)$ operation -- we implement TDML using a recursive approach, the Durbin–Levinson algorithm \cite{levinson1949wiener}.
        Specifically, for each time step $t$, the method computes the conditional expectation 
        $$
        \eta_t = E\bigl[y_t \mid y_1, \dots, y_{t-1}\bigr]
        $$
        and the corresponding prediction variance 
        $$
        v_t= \operatorname{Var}\bigl(y_t \mid y_1, \dots, y_{t-1}\bigr).
        $$ 

        These quantities are derived recursively using the theoretical autocovariance function of the process.

        The negative log-likelihood function is then formulated as
        $$
        \mathcal{L}(H, \mathbf{y}) = \frac{1}{2} \sum_{t=1}^{n} \left[ \log v_t + \frac{\bigl(y_t - \eta_t\bigr)^2}{v_t} \right],
        $$
        where the dependence on the Hurst exponent $H$ is implicit in the computation of both $\eta_t$ and $v_t$.
        The TDML estimator is obtained by minimizing the objective function:
        $$
        \hat{H}_\mathbf{y} := \arg\min_H \mathcal{L}(H, \mathbf{y}).
        $$

        \subsubsection{Whittle's method approximates the TDML method}
            For any stationary Gaussian sample the covariance matrix is Toeplitz. 
            By the Grenander–Szegő theorem a Toeplitz sequence is asymptotically equivalent to a circulant one \cite{grenander1958toeplitz}, and every circulant matrix is diagonalized by the discrete Fourier transform \cite{gray2006toeplitz}.
            Hence, after a discrete Fourier transform the time-domain maximum-likelihood log-determinant and quadratic form split into independent frequency terms, giving the Whittle likelihood objective \cite{whittle1951hypothesis}.
            The diagonalization error vanishes as the sample size grows, so the Whittle estimator has the same large-sample efficiency and normality as the exact time-domain maximum-likelihood estimator: Fox and Taqqu proved consistency and a central-limit theorem for long-range dependence \cite{fox1986large}, and Dahlhaus showed full asymptotic efficiency for self-similar Gaussian processes \cite{dahlhaus1989efficient}.
    
            The TDML approach with Durbin–Levinson recursion has the computational burden of $\mathcal{O}(n^2)$ in each optimization step, while Whittle's method needs to compute the theoretical spectral density at each step, and performs a single fast Fourier transform to calculate the periodogram at $\mathcal{O}(n \log n)$ cost.
            To achieve this great improvement in computational efficiency, Whittle's method trades exactness in finite samples, but still preserves full statistical efficiency asymptotically.
            
\section{Experiments}
    \subsection{General remarks}
        In the experiments below, unless stated otherwise, we test and compare the performance of different fBm Hurst parameter estimators.
        In general, we compare the estimators in different sequence length bins (e.g. $n=2^k,\,k\in \{7,8,\ldots ,15\}$).
        The fBm realizations were generated with $H\sim U(0,1)$, using the Davies-Harte method \cite{kroese2014spatial}.
        
        The inference times represent the estimated average computation time on one input sequence, and were calculated as: $t = w\cdot T/m$, where $m$ is the number of samples (input sequences) in a bin (e.g. $m=\num{100000}$), $w$ is the number of workers (processing threads), and $T$ is the total elapsed time for a given bin.
        Due to external factors beyond our experimental control -- such as the hardware configuration, overhead from multithreading, and shared server usage -- results presented should primarily be interpreted comparatively.

        For certain sequence length bins, in addition to the global Rooted Mean Square Error (RMSE) metric, we also calculate metrics such as RMSE, bias, and standard deviation locally in $H\in [h-dh, h+dh]$ windows, where $dh=0.05$ and $h\in\{0,0.001,0.002,\ldots,1\}$.

    \subsection{Approximations of the fGn spectrum}
\begin{figure}[H]
            \centering
            \includegraphics[width=0.48\textwidth]{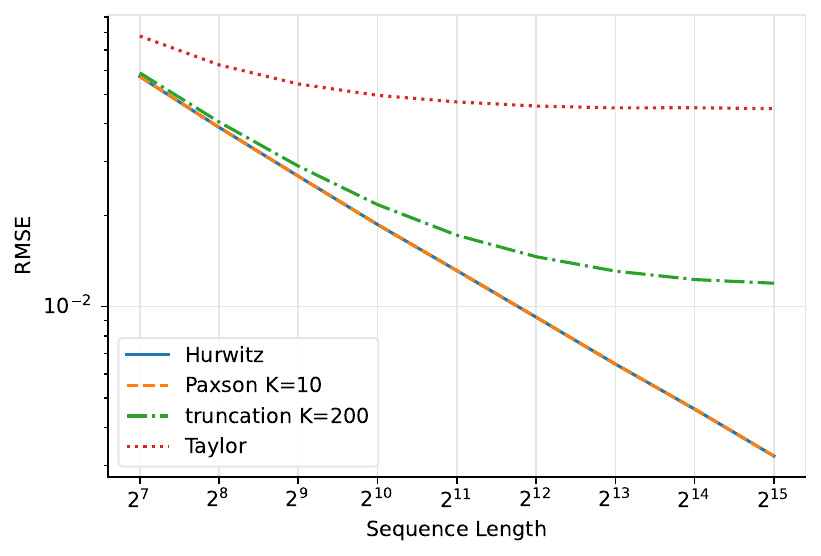}
            \vskip 2mm
            \caption{Empirical consistency of Whittle's method based on different approximations of the theoretical fGn spectrum.}
            \label{fig:fGn_spec_consistency_RMSE}
        \end{figure}

        In the following experiments, we compare the performance of Whittle's method based on the four different spectral density calculations discussed in Section~\ref{subsec_fgn_spect}.
        We compare the overall performance of these approaches across different input sequence lengths.
        The estimators based on the Hurwitz zeta reformulation and Paxson's corrected truncation of the fGn spectrum show remarkable consistency (Table~\ref{table:fGn_spec_consistency_RMSE}), evident from the linear trend in the log-log scale RMSE loss (Figure~\ref{fig:fGn_spec_consistency_RMSE}).
        In contrast, the uncorrected truncation and Taylor series-based approximations benefit much less significantly from longer input sequences.

\begin{figure}[H]
            \centering
            \includegraphics[width=0.48\textwidth]{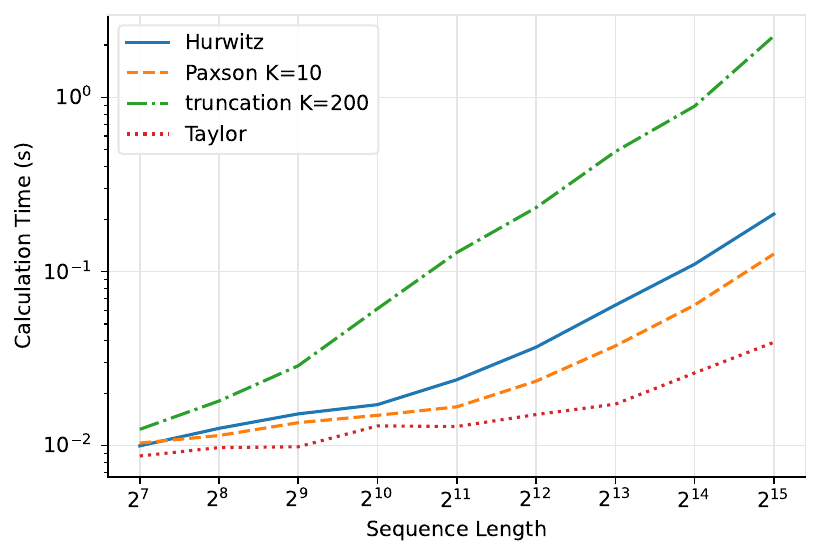}
            \vskip 2mm
            \caption{Per-sequence inference time comparison of Whittle's method based on different approximations of the fGn spectrum.}
            \label{fig:fGn_spec_consistency_time}
        \end{figure}

        Additionally, as shown in Figure~\ref{fig:fGn_spec_consistency_time}, both the Hurwitz zeta reformulation and Paxson's corrected truncation outperform the naïve truncation in terms of computational speed. 
        Although the Taylor series approximation is the fastest approach, its poor accuracy led us to select Paxson's corrected truncation as the default spectral density calculation for fGn in our implementation.

\begin{figure}[H]
             \centering
             \begin{subfigure}[b]{.48\textwidth}
                \centering
                \includegraphics[width=\textwidth]{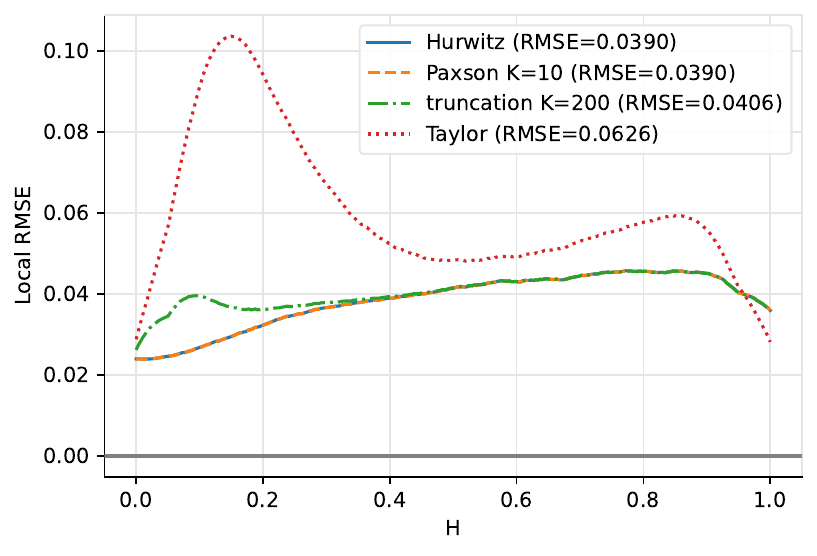}
                \subcaption{Sequence length $n=256$.}
                \vskip 5mm
                \label{fig:fGn_spec_local_RMSE_256}
             \end{subfigure}
             \hfill
             \begin{subfigure}[b]{.48\textwidth}
                \centering
                \includegraphics[width=\textwidth]{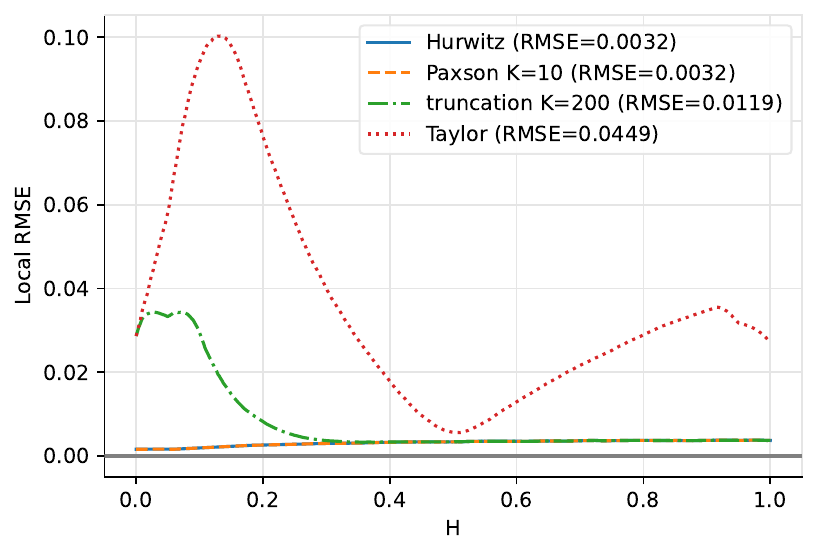}
                \subcaption{Sequence length $n=\num{32768}$.}
                \label{fig:fGn_spec_local_RMSE_32768}
             \end{subfigure}
             \vskip 5mm
             \caption{Local RMSE comparisons of different Whittle implementations by Hurst exponent value.}
            \label{fig:fGn_spec_local_RMSE}
        \end{figure}

        Figure~\ref{fig:fGn_spec_local_RMSE} compares the RMSE performance of the different approaches measured locally within small windows of $H$.
        Errors from the naïve truncation approach are predominantly localized in the Hurst range $H < 0.2$, making this method particularly ill-suited for certain practical applications \cite{volatility}.

\begin{figure}[H]
             \centering
             \hspace*{3.4mm}
             \begin{subfigure}[b]{.46\textwidth}
                \centering
                \includegraphics[width=\textwidth]{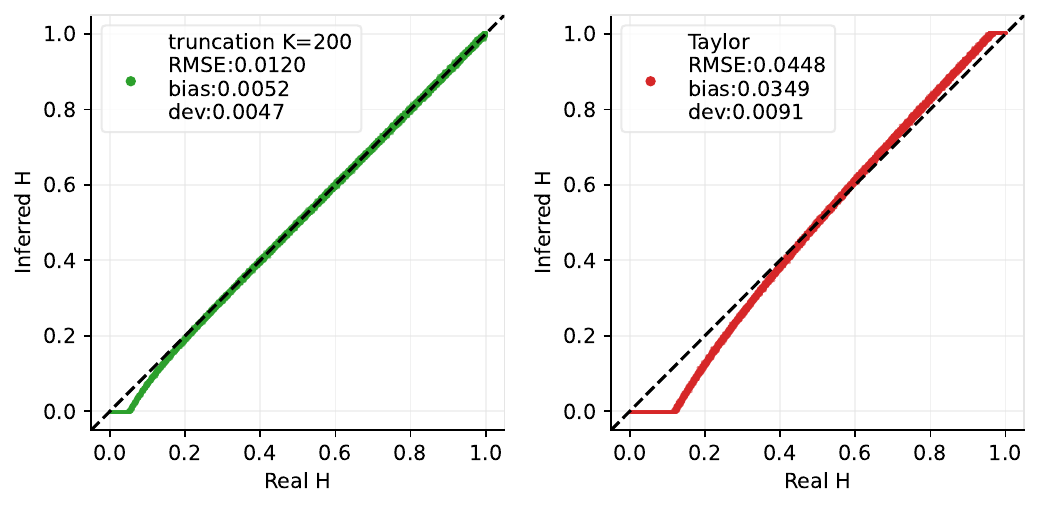}
                \label{fig:fGn_spec_bias_scatter}
             \end{subfigure}
             \hfill
             \begin{subfigure}[b]{.48\textwidth}
                \vskip -4mm
                \centering
                \includegraphics[width=\textwidth]{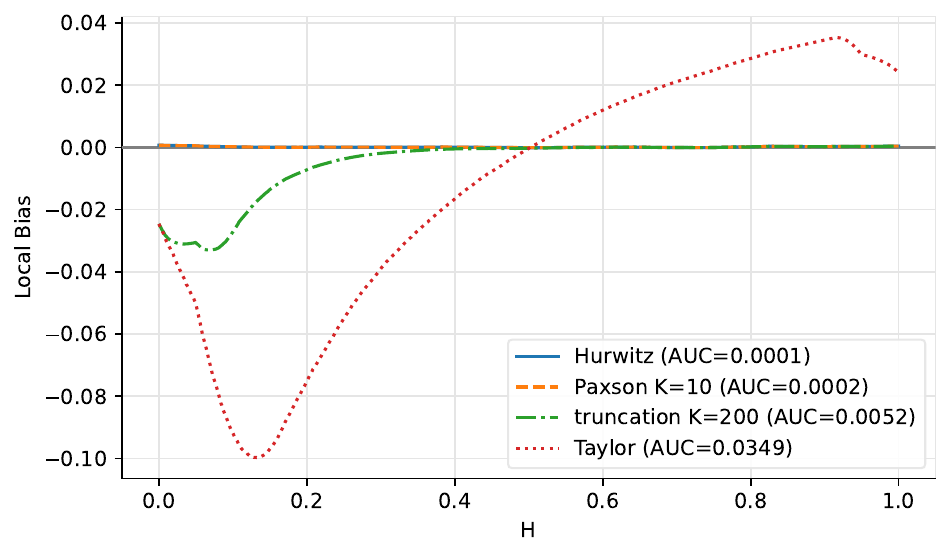}
                \label{fig:fGn_spec_bias_local}
             \end{subfigure}
             \vskip -3mm
             \caption{Scatter and bias comparisons of different Whittle implementations (sequence length $n=\num{32768}$).}
            \label{fig:fGn_spec_bias}
        \end{figure}

        Furthermore, localized errors from the naïve truncation and Taylor series approximations do not notably decrease with increasing input sequence lengths.
        This behavior can be explained by the inherent bias introduced by these approximations, as demonstrated in Figure~\ref{fig:fGn_spec_bias}.

    \subsection{Paxson's truncation}
\begin{figure}[H]
            \centering
            \includegraphics[width=0.48\textwidth]{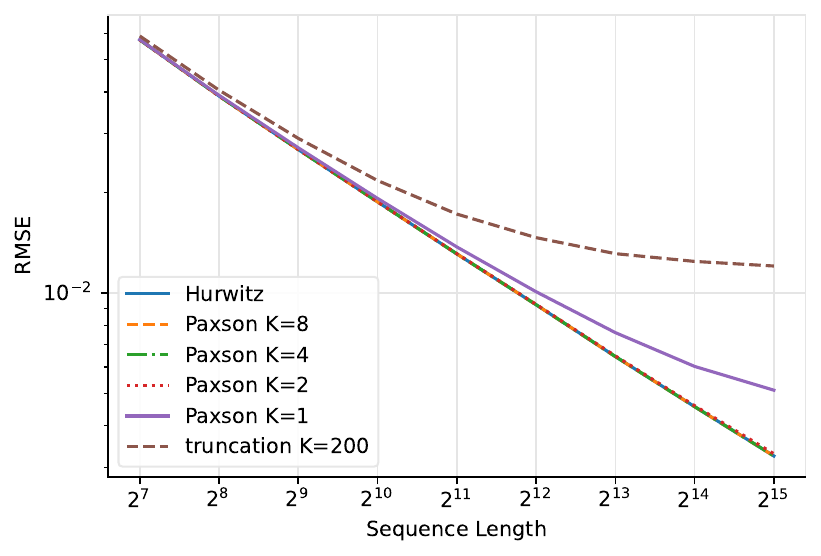}
            \vskip 2mm
            \caption{Empirical consistency of Whittle's method based on different Paxson truncations.}
            \label{fig:Paxson_consistency_RMSE}
        \end{figure}

\begin{figure}[H]
             \vskip -3mm
             \centering
             \begin{subfigure}[b]{.48\textwidth}
                \centering
                \includegraphics[width=\textwidth]{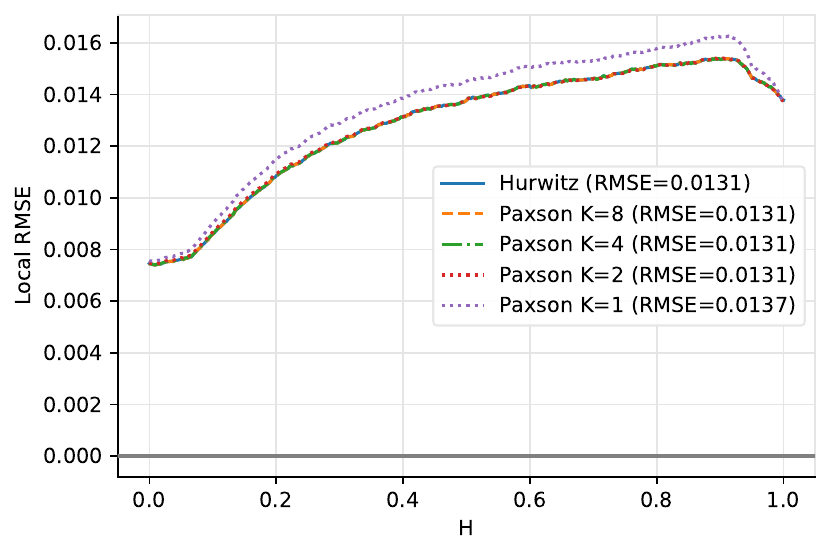}
                \label{fig:Paxson_local_RMSE}
             \end{subfigure}
             \hfill
             \begin{subfigure}[b]{.48\textwidth}
                \vskip -3mm
                \centering
                \includegraphics[width=\textwidth]{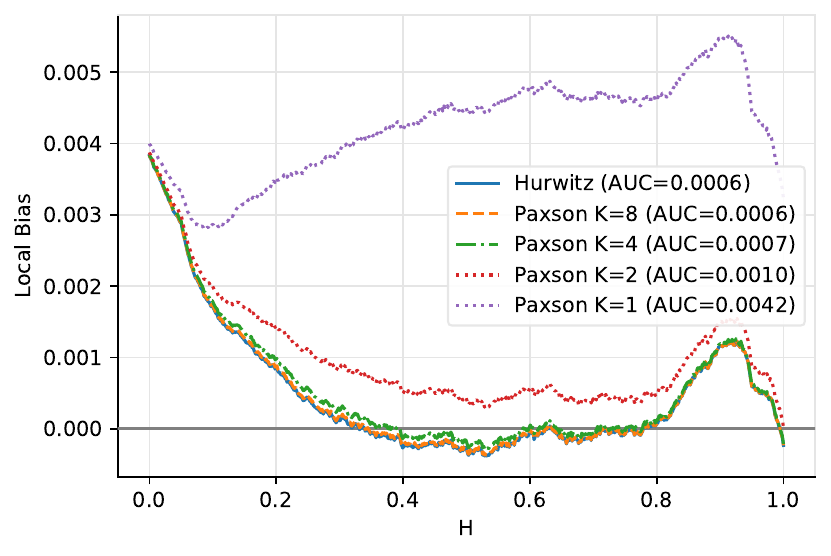}
                \label{fig:Paxson_local_biases}
             \end{subfigure}
             \vskip -3mm
             \caption{Local RMSE and bias of Whittle's method based on fGn spectrum with different Paxson truncations ($n=2048$).}
             \vskip 4mm
            \label{fig:Paxson_local}
        \end{figure}

        The amount of computation required for Paxson's truncation method directly depends on the number of terms ($K$).
        To practically determine how many terms are sufficient for accurate results, we evaluated the method using various values of $K$ until its performance was comparable to the method based on the Hurwitz zeta function.
        We observed that even with $K=1$, the approximation already yields remarkable results, and as $K$ increases further, the performance rapidly converges.
        By $K=8$, we have found virtually no difference between Paxson's truncation method and the approach using the Hurwitz zeta function (Figures~\ref{fig:Paxson_consistency_RMSE} and~\ref{fig:Paxson_local}).

        \begin{table}[H]
            \caption{Empirical consistency of Whittle’s Hurst estimation method based on different approximations of the theoretical fGn spectrum.}
            \vskip 3mm
            \centering
            \resizebox{0.48\textwidth}{!}{
            \begin{tabular}{lccccccc}
                & \multicolumn{7}{c}{$RMSE\;\;\left(\times 10^{-3}\right)$} \\
                \cmidrule(r{4pt}){2-8}
                $n$
                  & Hurwitz 
                  & $\stackrel{K=8}{\text{Paxson}}$
                  & $\stackrel{K=4}{\text{Paxson}}$
                  & $\stackrel{K=2}{\text{Paxson}}$
                  & $\stackrel{K=1}{\text{Paxson}}$
                  & $\stackrel{K=200}{\text{truncation}}$
                  & Taylor \\
                \midrule
                128   & 57.363 & 57.362 & 57.358 & 57.338 & 57.427 & 58.902 & 77.904 \\
                256   & 38.930 & 38.930 & 38.928 & 38.919 & 39.116 & 40.537 & 62.581 \\
                512   & 26.909 & 26.909 & 26.908 & 26.910 & 27.230 & 29.059 & 54.148 \\
                1024  & 18.751 & 18.751 & 18.751 & 18.758 & 19.205 & 21.723 & 49.662 \\
                2048  & 13.095 & 13.095 & 13.096 & 13.111 & 13.739 & 17.237 & 47.236 \\
                4096  & 9.2206 & 9.2208 & 9.2218 & 9.2424 & 10.085 & 14.648 & 45.792 \\
                8192  & 6.4481 & 6.4483 & 6.4497 & 6.4777 & 7.6051 & 13.115 & 45.142 \\
                16384 & 4.5623 & 4.5623 & 4.5630 & 4.5935 & 6.0231 & 12.431 & 45.208 \\
                32768 & 3.2457 & 3.2460 & 3.2478 & 3.2958 & 5.1119 & 12.043 & 44.876 \\
            \end{tabular}}
            \label{table:fGn_spec_consistency_RMSE}
        \end{table}

    \subsection{Other popular Hurst estimators}
        The performance of the Whittle estimator was benchmarked against several other widely-used methods.

        \textbf{Rescaled Range Analysis} ($R/S$) relates the cumulative range of a partial sum process to its standard deviation, yielding an exponent approximating $H$. 
        It is important for its historical context, because it was introduced alongside with the Hurst exponent \cite{hurst1951long}, but it is generally known to be less reliable than other methods \cite{taqqu1995estimators}.
        The $R/S$ method was imported from the \texttt{hurst} package~\cite{hurst}.

        \textbf{Higuchi’s method} estimates the \emph{fractal dimension} based on path roughness, employing grid-based box-counting \cite{higuchi1988approach}.
        As fractal dimension is directly related to the Hurst exponent \cite{falconer2013fractal}, this method provides a highly efficient way to estimate $H$.
        The method was imported from the \texttt{antropy} package~\cite{antropy}.

        \textbf{Detrended Fluctuation Analysis} (DFA) is a popular Hurst estimator robust for non-stationary processes  \cite{peng1994mosaic}.
        We imported this method from the \texttt{nolds} package~\cite{nolds}.
    
        \textbf{Variogram} of order $p=1$ (madogram) estimates the fractal dimension through log-log regression, leveraging local roughness captured in the moment domain \cite{mandelbrot1969robustness}.

\begin{figure}[H]
            \centering
            \includegraphics[width=0.48\textwidth]{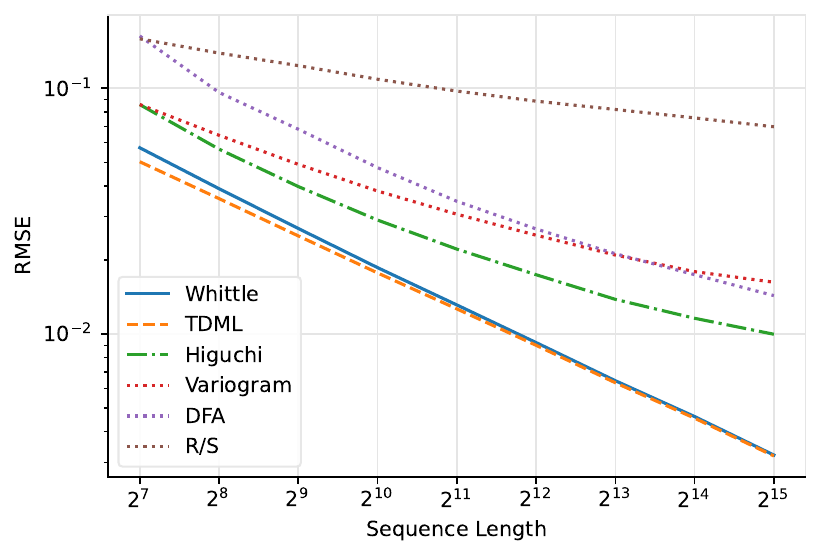}
            \vskip 2mm
            \caption{Empirical consistency of different Hurst estimators.}
            \label{fig:fBm_baselines_consistency_RMSE}
        \end{figure}

        As shown in Table~\ref{table:Bm_baselines_consistency_RMSE} and Figure~\ref{fig:fBm_baselines_consistency_RMSE}, our implementations of Whittle's and TDML methods consistently yield the most accurate estimates.
        We also observe that although TDML slightly outperforms Whittle's method for shorter input sequences, their performances converge as $n$ increases.

        \begin{table}[H]
            \vspace*{2mm}
            \caption{Empirical consistency comparison of different Hurst estimators. Whittle's method uses Paxson's truncation with $K=10$.}
            \vskip 3mm
            \centering
            \resizebox{0.48\textwidth}{!}{
            \begin{tabular}{lcccccc}
                & \multicolumn{6}{c}{$RMSE\;\;\left(\times 10^{-3}\right)$} \\
                \cmidrule(r{4pt}){2-7}
                $n$
                  & Whittle
                  & TDML
                  & Higuchi
                  & Variogram
                  & DFA
                  & R/S \\
                \midrule
                128   & 57.283 & 50.246 & 85.835 & 85.943 & 163.27 & 159.01 \\
                256   & 39.032 & 35.571 & 56.545 & 64.379 & 96.171 & 139.13 \\
                512   & 26.884 & 25.074 & 39.823 & 49.041 & 68.145 & 123.68 \\
                1024  & 18.627 & 17.724 & 29.092 & 38.186 & 47.543 & 108.87 \\
                2048  & 13.133 & 12.655 & 22.158 & 30.726 & 34.692 & 97.457 \\
                4096  & 9.2295 & 8.9977 & 17.432 & 25.241 & 26.775 & 88.649 \\
                8192  & 6.4547 & 6.3339 & 13.837 & 20.958 & 21.219 & 82.010 \\
                16384 & 4.6136 & 4.5437 & 11.589 & 17.942 & 17.436 & 75.789 \\
                32768 & 3.2162 & 3.1875 & 9.9766 & 16.244 & 14.331 & 69.737 \\
            \end{tabular}}
            \label{table:Bm_baselines_consistency_RMSE}
        \end{table}

        \begin{figure}[H]
            \centering
            \includegraphics[width=0.48\textwidth]{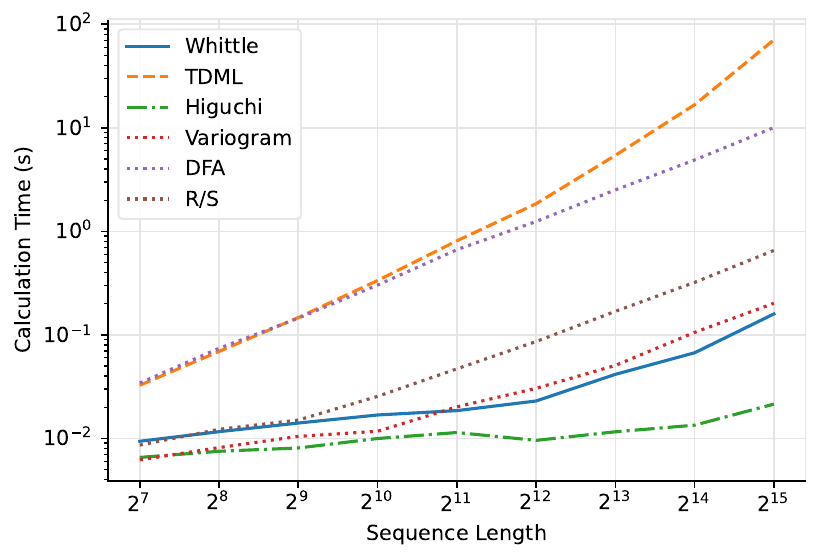}
            \vskip 2mm
            \caption{Computation time comparison of different Hurst estimators.}
            \label{fig:fBm_baselines_consistency_times}
        \end{figure}
        
        \begin{figure*}[h!]
             \centering
             \begin{subfigure}[b]{.53\textwidth}
                \centering
                \includegraphics[width=\textwidth]{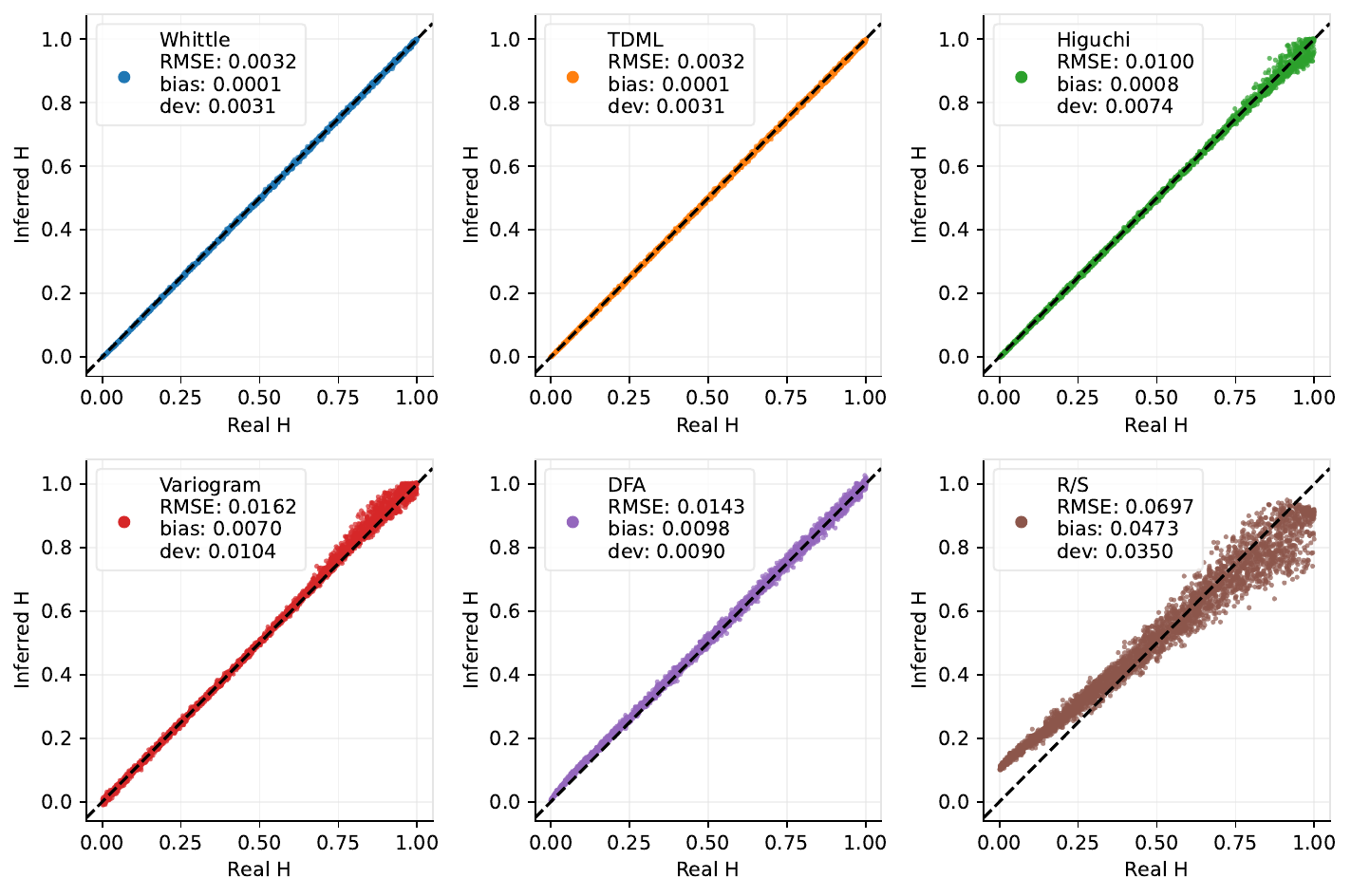}
                \vspace*{0.01mm}
                \vskip -1.3mm
                \subcaption{Scatter plots of different Hurst estimators ($n=\num{32768}$).}
                \label{fig:fBm_baselines_local_scatter}
             \end{subfigure}
             \hfill
             \begin{subfigure}[b]{.45\textwidth}
                \centering
                \includegraphics[width=\textwidth]{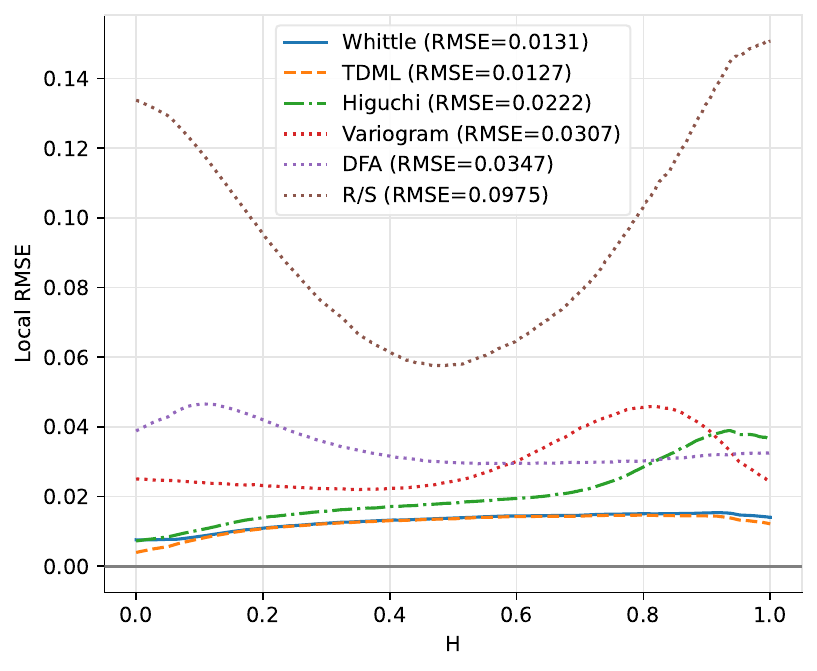}
                \subcaption{Local RMSE by Hurst exponent ($n=2048$).}
                \label{fig:fBm_baselines_local_RMSE}
             \end{subfigure}
             \vskip 4mm
             \caption{Scatter and RMSE plots of different Hurst estimators on fBm sequences.}
             \vskip 3mm
            \label{fig:fBm_baselines_local}
        \end{figure*}
        \newpage
        
        In terms of computational speed, Figure~\ref{fig:fBm_baselines_consistency_times} demonstrates that Whittle's method is surpassed only by Higuchi's algorithm.
        Furthermore, although TDML provides marginally lower errors, its computational complexity makes it one of the slowest methods evaluated.
        
\begin{figure}[H]
            \centering
            \includegraphics[width=0.48\textwidth]{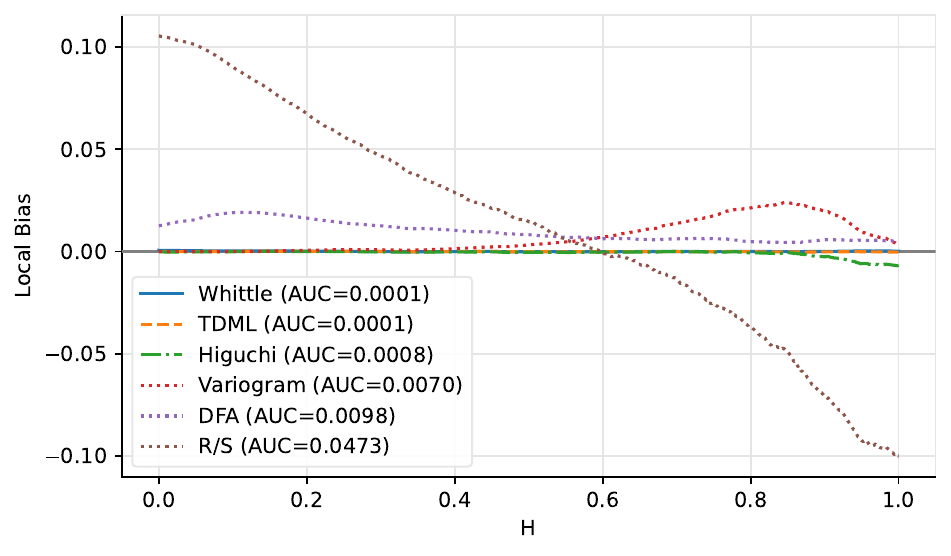}
            \caption{Bias of different Hurst estimators ($n=\num{32768}$).}
            \label{fig:fBm_baselines_bias}
        \end{figure}

\begin{figure}[H]
            \centering
            \vspace*{-2.8mm}
            \hspace*{1mm}
            \includegraphics[width=0.47\textwidth]{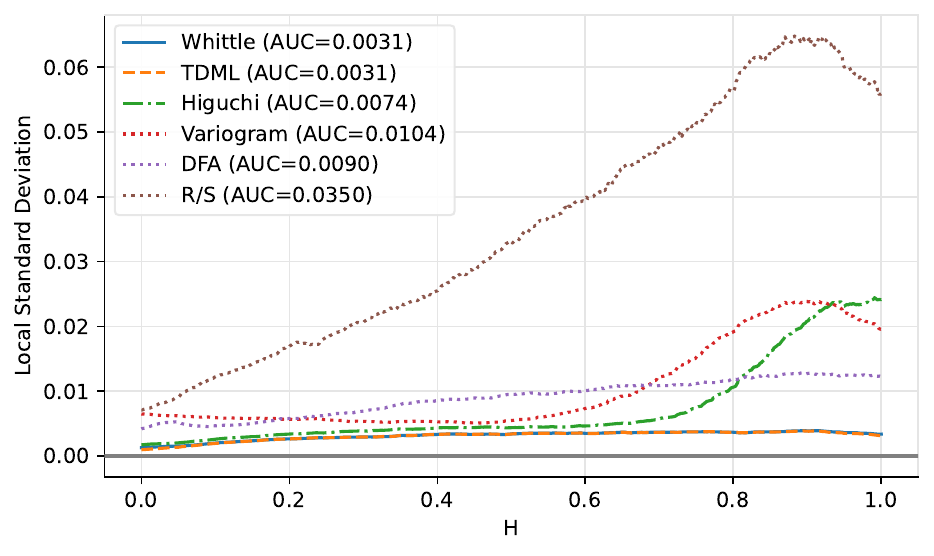}
            \caption{Deviation of different Hurst estimators ($n=\num{32768}$).}
            \label{fig:fBm_baselines_dev}
        \end{figure}

        Figure~\ref{fig:fBm_baselines_local} illustrates the local performance of the estimators.
        We observe that TDML and Whittle's method not only achieve the lowest overall RMSE but also provide the most accurate estimates consistently across the entire range of Hurst values.
        Additionally, Figure~\ref{fig:fBm_baselines_bias} highlights biases present in certain methods, such as DFA and $R/S$, and Figure~\ref{fig:fBm_baselines_dev} indicates significant deviations at specific Hurst values.
        For instance, Higuchi's method notably underperforms for $H > 0.7$.

    \subsection{ARFIMA} \label{ARFIMA}
\begin{figure}[h!]
            \centering
            \includegraphics[width=0.48\textwidth]{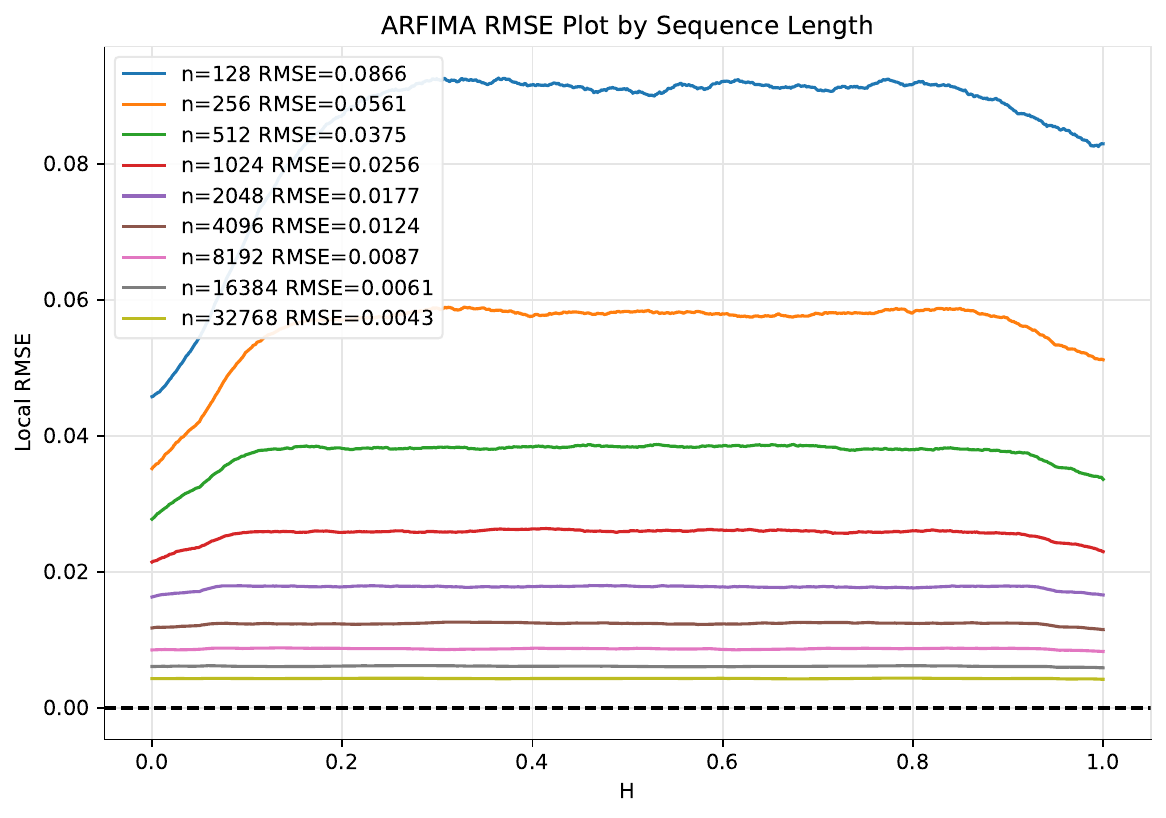}
            \caption{Local RMSE of Whittle's estimator based on the $\operatorname{ARFIMA}(0,H-0.5,0)$ spectrum by test sequence length.}
            \label{fig:ARFIMA}
         \end{figure}

        Figure~\ref{fig:ARFIMA} illustrates the convergence behavior of Whittle's method applied to $\operatorname{ARFIMA}(0, H - 0.5, 0)$ realizations, using the ARFIMA theoretical spectrum in the likelihood computation.
        Because the ARFIMA spectrum is much simpler to compute, the local error rate remains approximately constant across the parameter range, while the overall estimation error consistently decreases with increasing input sequence length.

    \subsection{Practical considerations on real-world data}
        \begin{figure*}[h!]
             \centering
             \begin{subfigure}[b]{.49\textwidth}
                \centering
                \includegraphics[width=\textwidth]{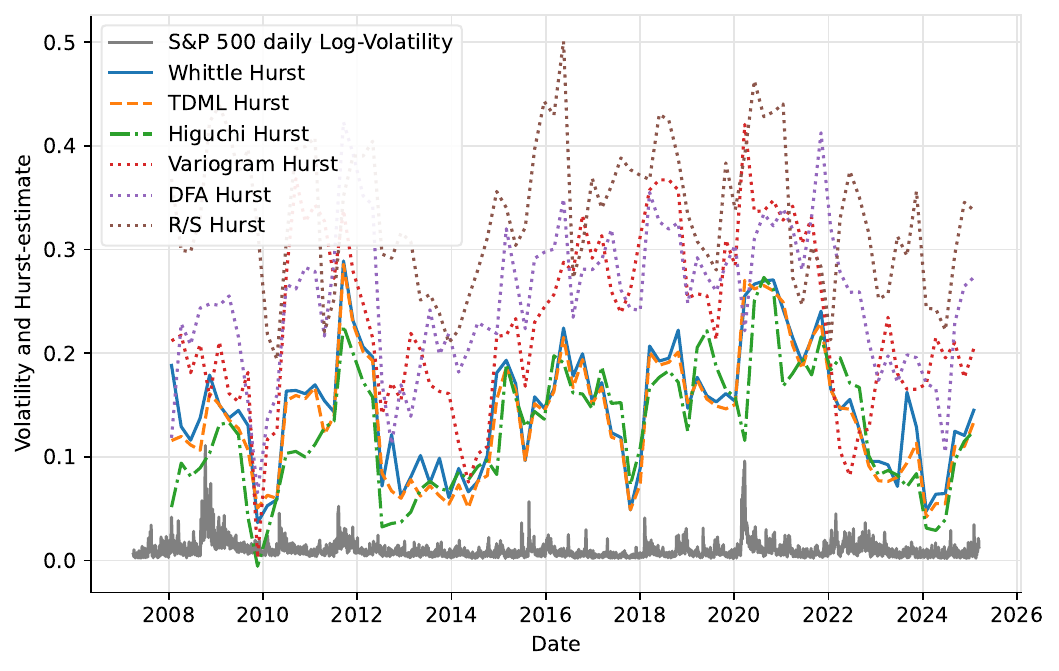}
                \subcaption{S\&P 500 log-volatility.}
                \label{fig:vol_SnP500}
             \end{subfigure}
             \hfill
             \begin{subfigure}[b]{.49\textwidth}
                \centering
                \includegraphics[width=\textwidth]{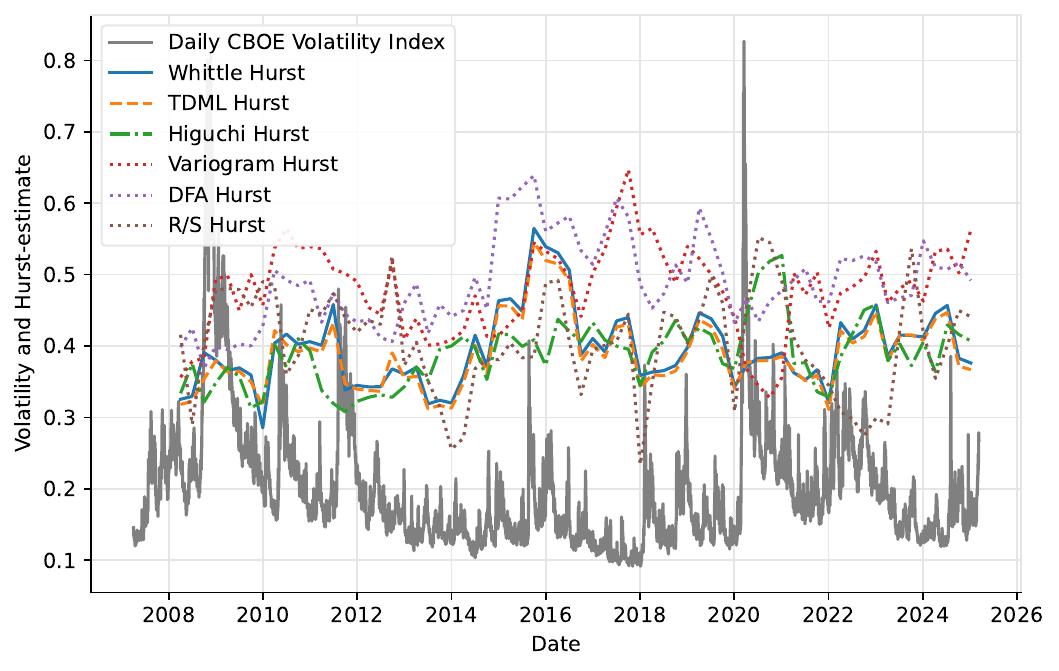}
                \subcaption{VIX index.}
                \label{fig:vol_VIX}
             \end{subfigure}
             \vskip 5mm
             \caption{Hurst-estimates on daily volatility data. Estimates use 252-day (one year) sliding windows with 189-day overlaps.}
             \vskip 3mm
            \label{fig:vol}
        \end{figure*}

        \begin{figure*}[h!]
             \vskip -1mm
             \centering
             \begin{subfigure}[b]{.49\textwidth}
                \centering
                \includegraphics[width=\textwidth]{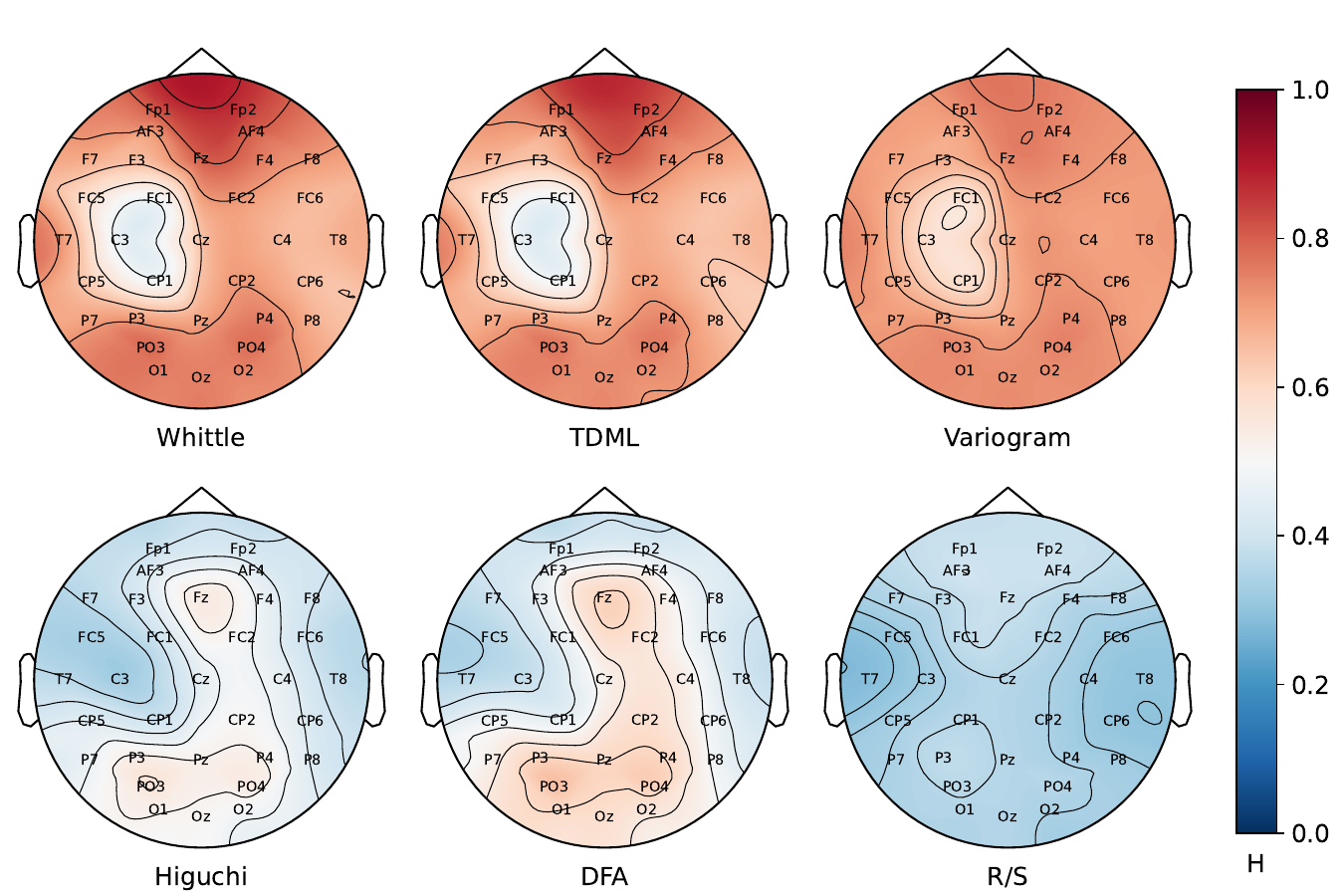}
                \subcaption{No band-pass filter.}
                \label{fig:ERP_fullspec}
             \end{subfigure}
             \hfill
             \begin{subfigure}[b]{.49\textwidth}
                \centering
                \includegraphics[width=\textwidth]{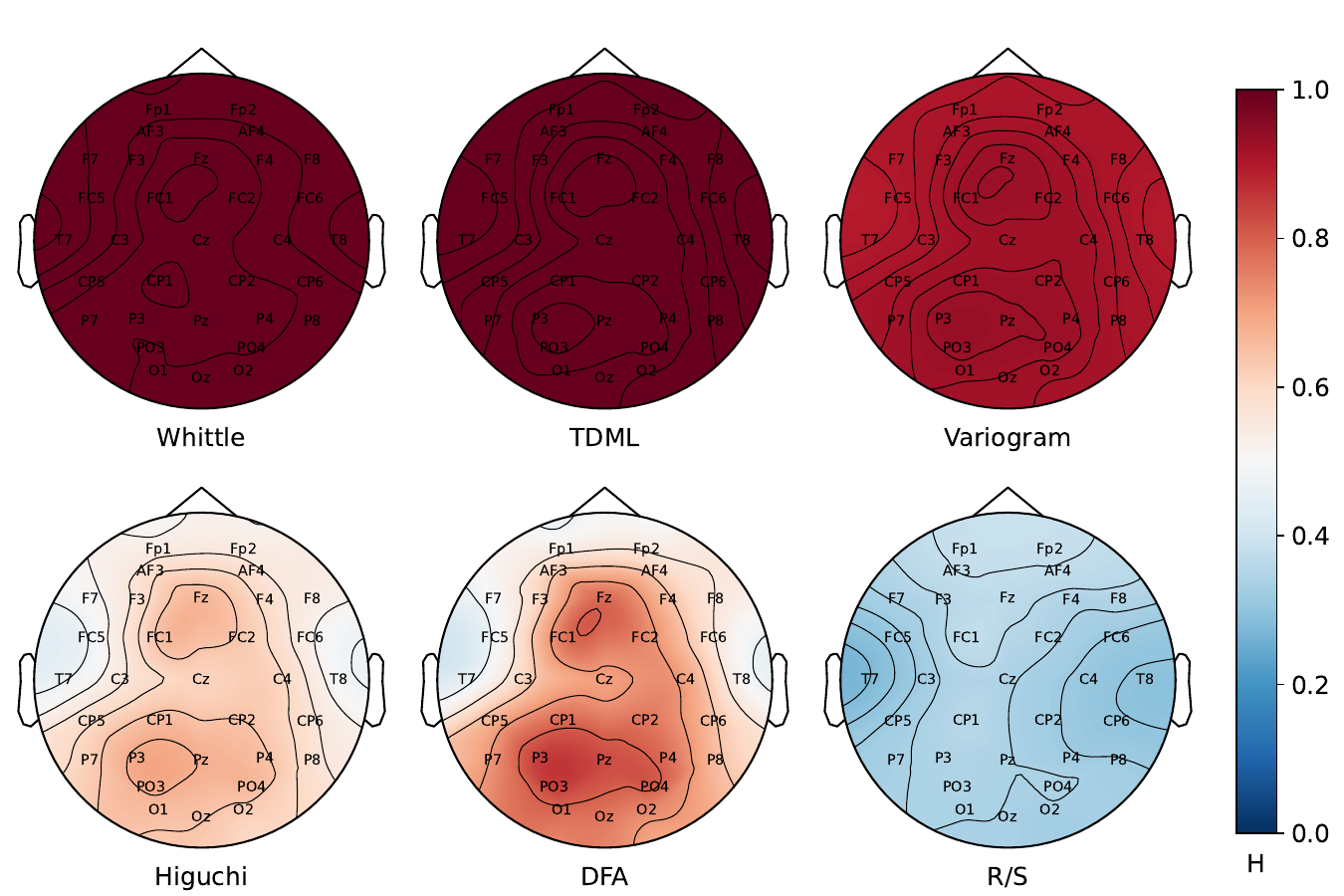}
                \subcaption{Band-pass filtered (0.1–\qty{80}{Hz}).}
                \label{fig:ERP_bandpass}
             \end{subfigure}
             \vskip 5mm
             \caption{Subject and event-wide average of ERP segment Hurst exponent estimations across EEG channels.}
             \vskip 3mm
            \label{fig:ERP}
        \end{figure*}
    
        Volatility in financial markets is often modeled by fBm with Hurst exponent $H\approx 0.1$~\cite{volatility}.
        Figure~\ref{fig:vol_SnP500} compares the different estimators on historical S\&P~500 log volatility calculated from 5-minute log-returns within non-overlapping daily windows.
        The results indicate that methods exhibiting a positive bias on synthetic data also tend to overestimate the Hurst exponent for this real-world rough data.
        
        A crucial practical consideration when estimating the Hurst exponent for volatility data is how volatility itself is computed.
        If volatility is calculated using overlapping time windows, artificial dependencies are introduced into the time series, increasing autocorrelation and consequently inflating Hurst exponent estimates.
        Figure~\ref{fig:vol_VIX} illustrates this effect using the daily Chicago Board Options Exchange (CBOE) Volatility Index, which measures the market's expectation of volatility over the next 30 days based on the S\&P~500.

        Estimations of the Hurst exponent also commonly arise in EEG data analysis.
        Figure~\ref{fig:ERP_fullspec} compares average Hurst exponent estimates by the different methods on EEG Event-Related Potential (ERP) segments \cite{luck2014introduction}.
        These ERP segments correspond to cuts in a short film and were extracted from EEG recordings of 15 subjects watching the same film.
        Hurst exponent estimations were performed individually for each EEG channel and segment, and the results were subsequently averaged across subjects and events.
        
        In EEG signal processing, specific preprocessing steps are routinely applied, particularly techniques affecting the signal's Fourier spectrum, such as notch and band-pass filtering.
        Notch filtering is commonly used to mitigate electrical grid noise (\qty{50}{Hz} in our case).
        Although our EEG signals were recorded at \qty{512}{Hz}, it is standard practice to apply a band-pass filter corresponding more closely to brain activity frequencies (e.g., 0.1–\qty{80}{Hz}).
        However, as demonstrated in Figure\ref{fig:ERP_bandpass}, this practice introduces complications in estimating the Hurst exponent, particularly for spectrum-based approaches such as Whittle's method.
        While this can be viewed as a limitation of the estimator, it also underscores an important consideration: when modeling real-world data with fractal-like processes such as fBm, we must remain aware of how preprocessing operations might fundamentally alter the fractal properties of the data and consider the sensitivity of our estimation methods to these changes.

\section{Conclusion}
    We have presented an in-depth discussion of Whittle's likelihood estimation method for the Hurst exponent, emphasizing practical implementation considerations, and performance in comparison with other estimation techniques.
    The method represents a compelling balance between computational efficiency and estimation accuracy.
    Its primary advantage lies in its frequency-domain formulation, which allows for rapid convergence even on large datasets.
    The various spectral density approximations offer flexibility in tailoring the estimator to different types of data, whether fGn or ARFIMA.
        
    However, the approach is not without limitations.
    The accuracy of the Whittle estimator depends on the quality of the spectral density approximation and assumptions on the periodogram.
    In scenarios where the spectral density is not approximated adequately by the theoretical model, the estimator will likely suffer from bias.
    On the other hand if we aim to model real-world data with a specific process such as fBm, the estimates provided by Whittle's method might be a better fit 
    compared to other methods.

\begin{ack}
    The research was supported by the Hungarian National Research, Development and Innovation Office within the framework of the Thematic Excellence Program 2021 -- National Research Sub programme: ``Artificial intelligence, large networks, data security: mathematical foundation and applications'' and the Artificial Intelligence National Laboratory Program (MILAB).
    We would also like to thank GitHub and neptune.ai for providing us academic access.
\end{ack}

\bibliography{whittlehurst}

\end{document}